\documentclass[aps,prb,reprint,amsmath,amssymb,floatfix,superscriptaddress]{revtex4-2}

\makeatletter
\let\frontmatter@footnote@produce\frontmatter@footnote@produce@endnote
\makeatother

\usepackage{float}
\usepackage{lipsum}
\usepackage{titlesec}
\usepackage{natbib}
\usepackage{graphicx}
\usepackage{physics}
\usepackage{bm}  
\usepackage{amsmath} 
\usepackage{braket} 
\usepackage[colorlinks,allcolors=blue,urlcolor=blue]{hyperref}

\begin{document}

\title{Magnetotransport in semiconductors and two-dimensional materials\\ from first principles}

\author{Dhruv C. Desai}
\affiliation{Department of Applied Physics and Materials Science, \protect\\ California Institute of Technology, Pasadena, California 91125}
\author{Bahdan Zviazhynski}
\affiliation{Trinity College, University of Cambridge, Cambridge, CB2 1TQ, United
Kingdom}
\author{Jin-Jian Zhou}
\affiliation{Department of Applied Physics and Materials Science, \protect\\ California Institute of Technology, Pasadena, California 91125}
\author{Marco Bernardi}
\email{bmarco@caltech.edu}
\affiliation{Department of Applied Physics and Materials Science, \protect\\ California Institute of Technology, Pasadena, California 91125}
%
%
\begin{abstract}
We demonstrate a first-principles method to study magnetotransport in materials by solving the Boltzmann transport equation (BTE) in the presence of an external magnetic field. Our approach employs \textit{ab initio} electron-phonon interactions and takes spin-orbit coupling into account. We apply our method to various semiconductors (Si and GaAs) and two-dimensional (2D) materials (graphene) as representative case studies. 
The magnetoresistance, Hall mobility and Hall factor in Si and GaAs are in very good agreement with experiments. In graphene, our method predicts a large magnetoresistance, consistent with experiments. Analysis of the steady-state electron occupations in graphene shows the dominant role of optical phonon scattering and the breaking of the relaxation time approximation. Our work provides a detailed understanding of the microscopic mechanisms governing magnetotransport coefficients, establishing the BTE in a magnetic field as a broadly applicable first-principles tool to investigate transport in semiconductors and 2D materials. 
\end{abstract} 
\maketitle
\titlespacing\section{-10pt}{04pt plus 2pt minus 2pt}{2pt plus 2pt minus 2pt}
\titlespacing\subsection{0pt}{12pt plus 4pt minus 2pt}{0pt plus 2pt minus 2pt}

\section{INTRODUCTION}
\vspace{5pt}
 Magnetic fields can strongly influence the electrical properties of materials, with changes quantified by magnetotransport coefficients such as the magnetoresistance (MR), Hall mobility and Hall factor~\cite{Ziman, Pippard}. In metals and semiconductors, the change in resistivity with magnetic field is typically small, but in certain semimetals, magnetic heterostructures and oxides the effects can be far greater or even dramatic, as in the case of giant and colossal MR~\cite{GMR,CMR}. Magnetotransport is of practical relevance for various applications, including sensors~\cite{Lenz1990}, magnetoresitive RAM and hard drives~\cite{Daughton1999}. In addition, measurements of the carrier concentration and electrical mobility require knowledge of the Hall factor. Therefore it is important to understand the physical mechanisms governing magnetotransport and develop methods to accurately predict the MR and Hall factor. 
 \\ 
 \indent
Experimentally, magnetotransport has been studied extensively in metals~\cite{Pippard} and simple semiconductors such as Si~\cite{Kirnas_Si_1974,Porter_Si_2011,Morin_Si_1954,Krag_Si_1959,Debye_Si_1954,Putley_1958} and GaAs~\cite{Rode_1975, Stillman_Gaas_1970, Blakemore_Gaas_1982}. 
More recently, measurements on two-dimensional materials have shown unconventional behaviors, such as large non-saturating MR at high fields in graphene~\cite{Matveev_Graphene_2018,Gopinadhan_Graphene_2013,Wang_Graphene_2014} and WTe$_2$~\cite{Ali_Wte2_2014}, and various studies have shown an interplay between band structure topology and magnetotransport, including the chiral anomaly and negative MR in topological semimetals~\cite{Xiong_Na3Bi_2015, Li_Cd3As2_2015, Armitage_Semimetal_2018}. These developments show that magnetotransport is a rapidly growing research arena.
\\
\indent
Early attempts to formulate theories of magnetotransport phenomena~\cite{Jan} focused on approximate solutions of the Boltzmann transport equation (BTE) in the relaxation time approximation (RTA)~\cite{Ziman}. Subsequent work using parametrized electronic band structures and electron-phonon ($e$-ph) interactions has shown calculations of the Hall factor in various materials~\cite{Lin_Si_1981, Reggiani_Diamond_1983}. 
Approaches beyond the RTA have also been proposed, for example by solving BTE in polar semiconductors in terms of infinite determinants~\cite{Lewis_Gaas_1955} or computing the phonon-limited Hall mobility in Si using deformation potential theory~\cite{Szmulowicz_Si_1983,Szmulowicz_Madarasz_Si_1983}. These models lack analytic closed-form solutions, and thus were implemented numerically, highlighting the need for computational approaches to study magnetotransport.
\\ 
\indent
In recent years, density functional theory (DFT)~\cite{Martin-book} and density functional perturbation theory (DFPT)~\cite{Baroni_DFPT_2001} have enabled \textit{ab initio} calculations of $e$-ph interactions. The resulting phonon-limited charge transport has been studied in various semiconductors and 2D materials in the framework of the BTE~\cite{Zhou2016, Jhalani2017, Lee_2018, Lee_2020, Jhalani-quad, Li2015, Chen2017, Li2018, Sohier2018}. 
First-principles studies of magnetotransport have lagged behind $-$ the only existing examples are two works by Macheda et al., who investigated an insulator (diamond)~\cite{Bonini-diamond} and very recently the Hall factor in graphene~\cite{Macheda_Graphene_2020}, using a conjugate gradient method to solve the BTE with a magnetic field. 
%
However, first-principles calculations of magnetotransport in semiconductors are still missing and the MR in 2D materials has not yet been computed. 
\\ 
\indent
Here we show calculations of the MR, Hall mobility and Hall factor, as a function of temperature and magnetic field, in  group-IV and polar semiconductors, focusing on the prototypical cases of Si and GaAs, and in graphene. Our approach, implemented in our {\sc{Perturbo}} code~\cite{perturbo2020}, solves the linearized BTE in a magnetic field using Jacobi iteration to obtain the conductivity tensor and from it the magnetotransport properties. The calculations employ \textit{ab initio} $e$-ph interactions and include spin-orbit coupling (SOC), which is particularly important for holes. 
Extensive comparisons with experiments demonstrate the accuracy of our first-principles magnetotransport calculations for semiconductors. Analysis of the relative occupation changes in momentum space shows the dominant role of backscattering due to optical phonons and the breaking of the RTA in graphene.  
Taken together, our work demonstrates an accurate method to investigate magnetotransport in semiconductors and 2D materials and clarify the underlying microscopic mechanisms.\\

\vspace{-15pt}
\section{METHODS}
\vspace{0.1cm}
%
%
\subsection{Magnetotransport properties and BTE}
\vspace{0.1cm}
In the presence of small electric ($\mathbf{E}$) and magnetic ($\mathbf{B}$) fields, the current density $\boldsymbol{\mathrm{J}}$ can be written as
\begin{equation} \label{eq:ohmslaw}
    J_{i} =\sum_{j=1}^3 \sigma_{ij}(\mathbf{B})E_{j}
\end{equation}
with the conductivity tensor $\sigma_{ij}$ expanded as~\cite{Ziman}
\begin{equation} \label{taylorexpcond}
\sigma_{ij}(\mathbf{B}) = \sigma_{ij}^{(0)} + 
\sigma_{ijk}^{(1)}B_{k} + 
\sigma_{ijkl}^{(2)}B_{k}B_{l} + \ldots
\end{equation}
with implied summations over repeated indices (which correspond to Cartesian components). We write the current in terms of electronic occupations $f_{n\mathbf{k}}$ and band velocities $v_{n\mathbf{k}}$ ($n$ is the band index and $\mathbf{k}$ the crystal momentum of the electronic state):
\begin{equation}
\label{currentdefinition}
    \boldsymbol{\mathrm{J}} = \frac{-Se}{\mathcal{N}_{\mathbf{k}} \Omega} \sum_{n\mathbf{k}}f_{n\mathbf{k}}\mathbf{v}_{n\mathbf{k}},
\end{equation}
where $e$ is the absolute value of the electric charge, $S$ the spin degeneracy, $\mathcal{N}_{\mathbf{k}}$ the number of unit cells, and $\Omega$ their volume. 
At steady-state, the BTE in the presence of both electric and magnetic fields reads~\cite{Ziman}
\begin{equation} \label{eq:full_BTE}
    e\frac{\partial f_{n \mathbf{k}}}{\partial \epsilon_{n \mathbf{k}}} \mathbf{v}_{n \mathbf{k}} \cdot \mathbf{E} + \frac{e}{\hbar} (\mathbf{v}_{n\mathbf{k}} \cross \mathbf{B}) \cdot \nabla_{\mathbf{k}}f_{n\mathbf{k}} + \mathcal{I}^{\rm{e-ph}}[f_{n\mathbf{k}}] = 0
\end{equation}
where $\epsilon_{n\mathbf{k}}$ are electronic energies, and the last term includes $e$-ph collision processes consisting of absorption or emission of a phonon~\cite{perturbo2020}. Expanding $f_{n\mathbf{k}}$ to leading order in $\mathbf{E}$, we write  $f_{n\mathbf{k}} - f_{n\mathbf{k}}^0 = -f_{n\mathbf{k}}^0(1 - f_{n\mathbf{k}}^0)\frac{e\mathbf{E}}{k_{B}T} \cdot \mathbf{F}_{n\mathbf{k}}$, and solve for the unknown occupation changes $\mathbf{F}_{n\mathbf{k}}$~\cite{perturbo2020}. Factoring out $-e\mathbf{E}f_{n\mathbf{k}}^0(1 - f_{n\mathbf{k}}^0)/k_{B}T $, we obtain the linearized BTE 
%
\begin{equation} \label{eq:Linearized_BTE}
\begin{split}
\mathbf{v}_{n\mathbf{k}}+ \frac{e}{\hbar}
(\mathbf{v}_{n\mathbf{k}} \cross \mathbf{B})\nabla_{\mathbf{k}}\mathbf{F}_{n\mathbf{k}} = \\
\frac{1}{\mathcal{N}_{\mathbf{q}}} \sum_{m,\nu \mathbf{q}} W_{n\mathbf{k},m\mathbf{k+q}}^{\nu \mathbf{q}} (\mathbf{F}_{n\mathbf{k}} - \mathbf{F}_{m\mathbf{k+q}}) ,
\end{split}
\end{equation}
where $\nu$ is the phonon mode index, $\mathbf{q}$ the phonon wavevector and $\mathcal{N}_{\mathbf{q}}$ the number of $\mathbf{q}$ points used in the summation. Here,  $W_{n\mathbf{k},m\mathbf{k+q}}^{\nu \mathbf{q}}$ is the scattering rate from $\ket{n\mathbf{k}}$ to $\ket{m\mathbf{k+q}}$ and takes into account both phonon absorption and emission processes~\cite{perturbo2020}. 
We solve for $\mathbf{F}_{n\mathbf{k}}$ by rearranging terms in Eq.~(\ref{eq:Linearized_BTE}) and using the iterative Jacobi scheme. 
For each iteration $i$, we get
%
%
\begin{equation} \label{eq:Linearized_BTE_Bfield}
\begin{split}
      \mathbf{F}_{n\mathbf{k}}^{(i+1)} = \mathbf{v}_{n\mathbf{k}} \tau_{n\mathbf{k}} + \frac{\tau_{n\mathbf{k}}}{\mathcal{N}_{\mathbf{q}}} \sum_{m,\nu \mathbf{q}} W_{n\mathbf{k},m\mathbf{k+q}}^{\nu \mathbf{q}} \mathbf{F}_{m \mathbf{k+q}}^{(i)}  \\ + \frac{e}{\hbar} \tau_{n\mathbf{k}} (\mathbf{v}_{n\mathbf{k}} \cross \mathbf{B})\nabla_{\mathbf{k}}\mathbf{F}_{n\mathbf{k}}^{(i)},
\end{split}
\end{equation}
where $\mathrm{\tau_{n\mathbf{k}}}$ is the relaxation time. The term containing the gradient in $\mathbf{k}$, $\nabla_{\mathbf{k}}\mathbf{F}_{n\mathbf{k}}$, is computed using the central finite difference approximation in Ref.~\cite{Mostofi_wannier90_2008}.
Starting with the RTA solution as the initial guess, $\mathbf{F}_{n\mathbf{k}}=\mathbf{v}_{n\mathbf{k}}\tau_{n\mathbf{k}}$, we evaluate the right-hand side of Eq.~(\ref{eq:Linearized_BTE_Bfield}) to update the solution $\mathbf{F}_{n\mathbf{k}}$, iterating this procedure until convergence.
\\
\indent
%
%
Expanding $f_{n \mathbf{k}}$ in Eq.~(\ref{currentdefinition}), we obtain~\cite{perturbo2020}
%
%
\begin{equation} \label{eq:Conductivity_tensor}
    \sigma_{ij} = \frac{e^2S}{\mathcal{N}_{\mathbf{k}}\Omega k_{\rm{B}}T} \sum_{n\mathbf{k}} f_{n\mathbf{k}}^{0}(1-f_{n \mathbf{k}}^{0})\, (\mathbf{v}_{n\mathbf{k}})_i  (\mathbf{F}_{n\mathbf{k}})_j,
\end{equation}
We can calculate the magnetotransport coefficients from this conductivity tensor because of its implicit dependence on $\mathbf{B}$ through Eq.~(\ref{eq:Linearized_BTE_Bfield}). The MR can be obtained from the resistivity tensor $\rho(\mathbf{B})$ = $\sigma^{-1}(\mathbf{B})$ using~\cite{Ziman}:
\begin{equation}
\mathrm{
    MR }= \frac{\rho(\mathbf{B}) - \rho(0)}{\rho(0)}.
\end{equation}
At low fields, the MR is expected to be quadratic in the magnetic field~\cite{Jones_1973}.
In most materials, the MR perpendicular to $\mathbf{B}$ (transverse MR) is small and positive $-$ classically, this increase in resistivity can be viewed as a result of the Lorentz force deviating charge carriers from their initial trajectories.
\\
\indent
First-principles calculations typically compute the drift mobility $\mu_{\rm{d}}$ in zero magnetic field, whereas in experiments a common practice is to obtain the mobility from Hall measurements~\cite{Sze_2006}; the resulting Hall mobility is defined as $\mu_{H}$ = $\sigma_{\rm{d}} R_{\rm{H}}$, where $\sigma_{\rm{d}}$ is the drift conductivity and $R_{\rm{H}}$ the Hall coefficient. In Drude theory, $R_{\rm{H}}$ evaluates to $1/ne$ for a carrier concentration $n$~\cite{Ziman}, so $\mu_{\rm{H}}$ = $\mu_{\rm{d}}$. However, when the dependence of the relaxation time on electronic state is taken into account, $R_{\rm{H}}$ deviates from the Drude value by the Hall factor $r$ = $\mu_{\rm{H}}/\mu_{\rm{d}}$~\cite{Lin_Si_1981}, so the Hall and drift mobilities differ by the Hall factor. For systems with cubic symmetry and $\mathbf{B}$ field in the $z$ direction, the Hall factor is $r$ = $n e\, \sigma_{xyz}^{(1)}/ \sigma_{xx}^{(0)}$~\cite{Reggiani_Diamond_1983}. \\

\begin{figure*}[t]
\centering 
\includegraphics[width=0.85\textwidth]{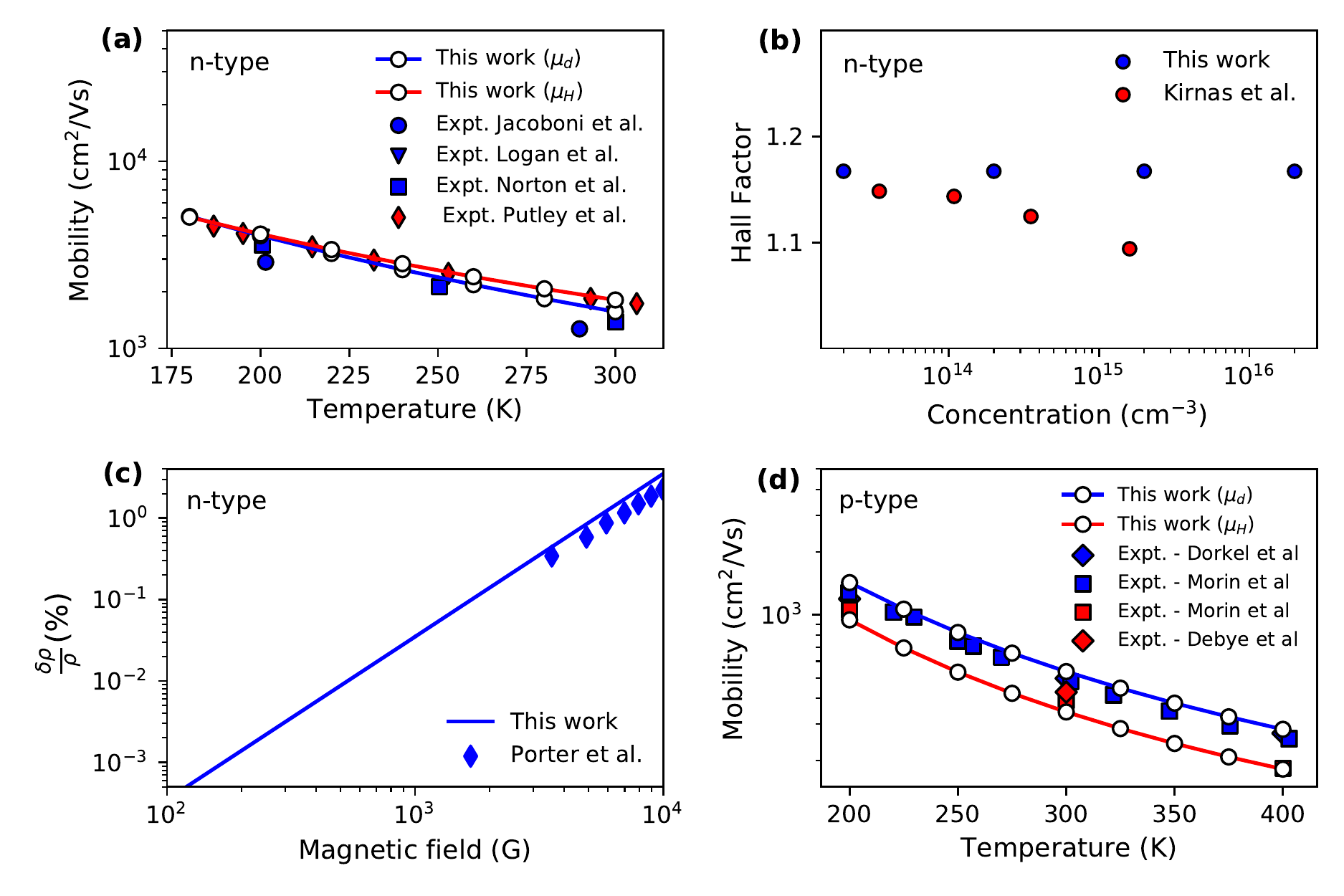}
\vspace{-15pt}
\caption{(a) Drift and Hall mobilities, in blue and red respectively, as a function of temperature in n-type silicon (experimental data are from Refs.~\cite{Jacoboni_Si_1977, Logan_Si_1960,Norton_Si_1973,Putley_1958}). 
(b) Hall factor at $300$~K as a function of carrier concentration in n-type silicon. 
(c) Transverse MR as a function of magnetic field in n-type silicon, compared with experiments from Ref.~\cite{Porter_Si_2011}.
(d) Drift and Hall mobilities as a function of temperature in p-type silicon (experimental data are taken from Refs.~\cite{Morin_Si_1954, Debye_Si_1954,Dorkel_Si_1981}). 
}\label{fig:silicon_data}
\end{figure*}
\subsection{Computational details}
\vspace{5pt}
We apply our approach to Si, GaAs and graphene. Their ground state is computed using DFT in the local density approximation, with a plane-wave basis set and norm-conserving pseudopotentials, using the {\sc{Quantum Espresso}} package. We use plane-wave kinetic energy cutoffs of 40 Ry for Si, 72 Ry for GaAs and 90 Ry for graphene and relaxed lattice parameters of 5.43~$\mathrm{\AA}$ for Si,  5.56~$\mathrm{\AA}$ for GaAs, and 2.44~$\mathrm{\AA}$ for graphene. The phonon dispersions and $e$-ph perturbation potentials on coarse $\mathbf{q}$-point grids are computed with DFPT~\cite{Giannozzi_QE_2009} and the Wannier functions are obtained using {\sc{Wannier90}}~\cite{Mostofi_2014}. We use coarse electron $\mathbf{k}$-point and phonon $\mathbf{q}$-point grids of $8\times8\times8$ for Si and GaAs and $36\times36\times1$ $\mathbf{k}$- and $18\times18\times1$ $\mathbf{q}$-points for graphene. We compute and interpolate the $e$-ph matrix elements using our {\sc{Perturbo}} open source package~\cite{perturbo2020}. Calculations with SOC~\cite{perturbo2020,Jinsoo} employ fully relativistic pseudopotentials. 
We implement the iterative solution of the BTE in a magnetic field in {\sc{Perturbo}}, and use very fine, equal and uniform $\mathbf{k}$- and $\mathbf{q}$-point grids (with $140^3$ points for Si, $650^3$ for GaAs and $1800^2$ for graphene) to converge the BTE solutions. The conductivity tensor is obtained via tetrahedron integration~\cite{perturbo2020}.

\section{RESULTS}
\vspace{0.1cm}
\subsection{Silicon}
We compute the drift and Hall mobilities, Hall factor and MR as a function of temperature for Si, and compare the computed results with experiments.  Figure~\ref{fig:silicon_data}(a) shows the Hall and drift electron mobilities in n-type silicon. The agreement with experimental data from Refs.~\cite{Jacoboni_Si_1977, Logan_Si_1960,Norton_Si_1973,Putley_1958} is excellent. As expected for electron carriers, the Hall mobility is greater than the drift mobility at all temperatures.
The computed Hall factor, $r = \mu_{\rm{H}} / \mu_{\rm{d}}$, increases slightly with temperature, as evidenced by higher deviations between $\mu_{\rm{H}}$ and $\mu_{\rm{d}}$ for higher temperatures. 
\\
\indent
The Hall factor for electrons is shown in Fig.~ \ref{fig:silicon_data}(b) at 300~K as a function of carrier concentration, which can be tuned in our calculations by changing the chemical potential. At low carrier density, our computed Hall factor is very close to the accepted value of $\sim$1.15 in n-type Si~\cite{Kirnas_Si_1974}. The computed Hall factor is within $\sim$10\% of experiment at all carrier concentrations, a noteworthy result for a calculation without adjustable parameters. 
We attribute the increasing deviation from experiments at higher concentrations to scattering from ionized impurities not taken into account in this work.
\\
\indent
The transverse MR is a common figure of merit for various applications. In Fig.~\ref{fig:silicon_data}(c), we plot the transverse MR as a function of magnetic field for electron carriers in n-type Si. 
The computed MR is in very good agreement with experiments from Ref.~\cite{Porter_Si_2011}. 
The results are computed directly in the low field regime $\mu_{\rm{H}}\mathrm{B} \ll 1$, but  calculations at higher fields ($B> 2\cdot 10^3$ G) did not converge under the Jacobi scheme $-$ the linearized BTE approximation is adequate only at low field $-$ and were extrapolated using a parabolic dependence. %
Remaining differences between experiment and theory may be due to various factors, including uncertainty in the experimental temperature and doping concentration, as well as inevitable small deviations from experiment of the computed band structure and phonon dispersions. 
\\
\indent
Figure \ref{fig:silicon_data}(d) shows the computed Hall and drift mobilities of hole carriers and compares them with experimental data for p-type silicon. For hole carriers, we find that calculations without SOC fail to produce an isotropic conductivity tensor, a key sanity check for Si (for electrons, SOC has only a minor effect). 
Including SOC in our band structure and $e$-ph calculations~\cite{perturbo2020,Jinsoo} is key to obtaining isotropic magnetotransport for holes. 
The hole mobilities are in very good agreement with data for p-type Si~\cite{Dorkel_Si_1981,Morin_Si_1954,Debye_Si_1954}.
For hole carriers, correctly, we obtain a behavior opposite to electrons, $\mu_{\rm{H}} \!<\! \mu_{\rm{d}}$ in the entire temperature range and thus a Hall factor $r\!<\!1$. 
Our computed low field MR coefficient, $\rm{MR}/B^2$, is $\mathrm{6.46 \cdot 10^{5}}$ $\mathrm{cm^4/V^2s^2}$ for holes at 300~K,  within 10\% of the measured value of $\mathrm{5.90 \cdot 10^{5}}$ $\mathrm{cm^4/V^2s^2}$~\cite{Long_Si_1958}. These results show that including SOC makes accurate magnetotransport calculations possible for hole carriers in semiconductors.
\vspace{-10pt}
%
%
\begin{figure}[t]
\centering
\includegraphics[width=0.94\columnwidth ]{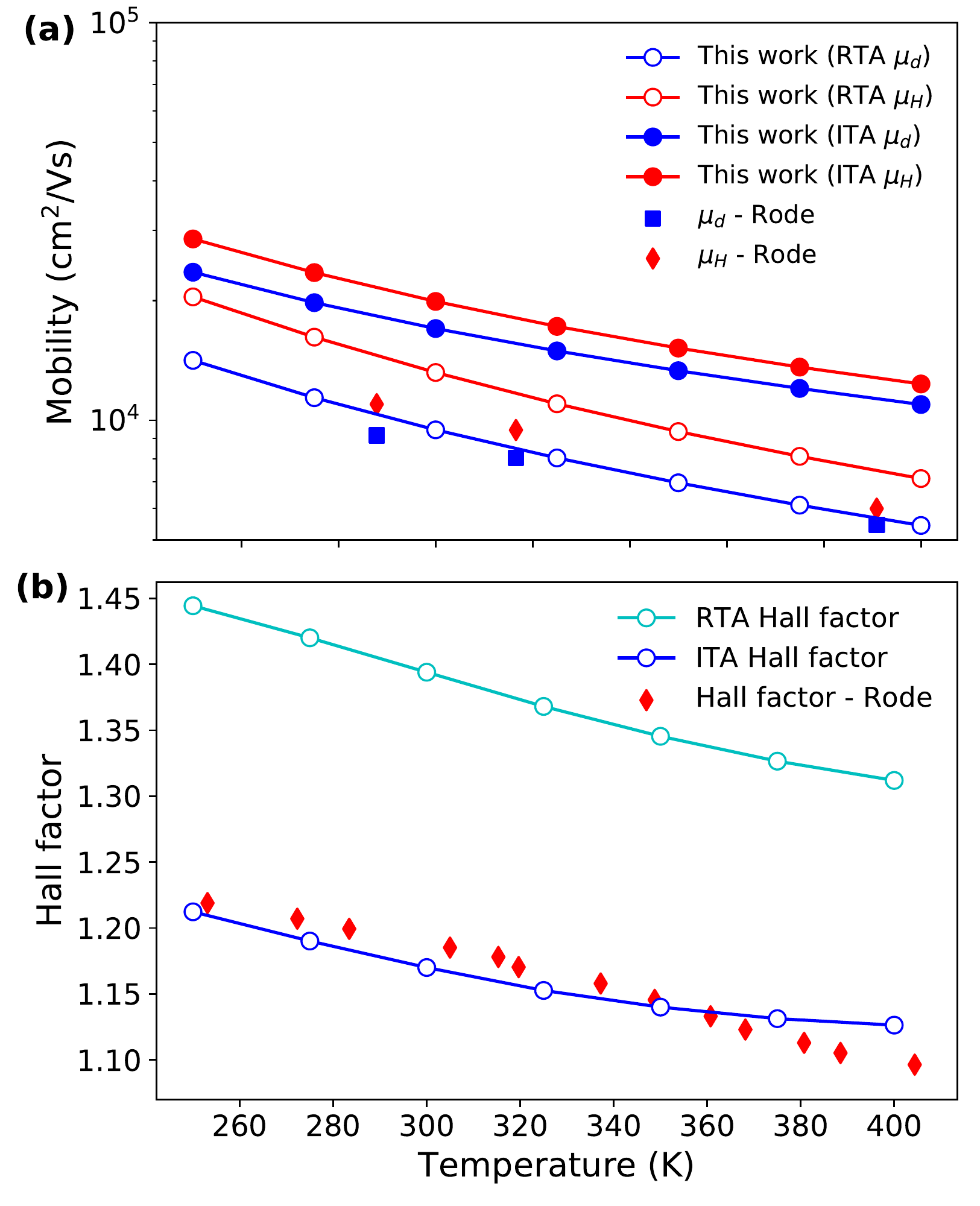}
\vspace{-10pt}
\caption{(a) Drift and Hall mobilities in GaAs as a function of temperature for an electron concentration $n=\mathrm{10^{16} \,cm^{-3}}$. (b) Hall factor vs temperature for electrons in GaAs.
}\label{fig:gaas}
\end{figure}
\vspace{-30pt}
\subsection{Gallium arsenide}
The drift mobility has been studied extensively from first principles in GaAs~\cite{Zhou2016,Chen2017,Lee_2020}. Due to its polar character, electrons in GaAs couple strongly with longitudinal optical (LO) phonons through the Fr\"ohlich interaction. We have recently shown that the iterative solution of the linearized BTE (ITA in short) overestimates the mobility and that including electron-two-phonon ($e$-2ph) scattering processes significantly improves the result; the RTA also gives a mobility in agreement with experiment~\cite{Zhou2016}, but due to compensation of errors~\cite{Lee_2020}. We find that the same trends also hold for the Hall mobility. Figure \ref{fig:gaas}(a) shows the drift and Hall mobilities for electrons in GaAs as a function of temperature. The experimental Hall mobility shown for comparison is obtained as $\mu_{\rm{H}}$ = $\mu_{\rm{d}} r$ with values of $\mu_{\rm{d}}$ and $r$ from  Ref.~\cite{Rode_1975}.
\\
\indent
The ITA overestimates both the drift and Hall mobilities, by a factor of $\sim$2 at 300~K, while the RTA is in better agreement with experiments due to error compensation~\cite{Lee_2020}.  
The Hall factor $r = \mu_{\rm{H}}/\mu_{\rm{d}}$ for both approaches is correctly greater than 1, but the Hall factor for ITA is much closer to the measured data [Fig.~\ref{fig:gaas}(b)]. Although each of the Hall and drift mobilities are overestimated in the ITA, their ratio is predicted accurately; we cannot establish whether this result is a coincidence or due to cancellation of effects from $e$-2ph processes in the ratio. Overall, these trends show that for polar semiconductors first-principles magnetotransport calculations have an accuracy similar to calculations without magnetic field.

\subsection{Graphene}
\begin{figure}[!hb]
\centering 
\includegraphics[width=0.95\columnwidth]{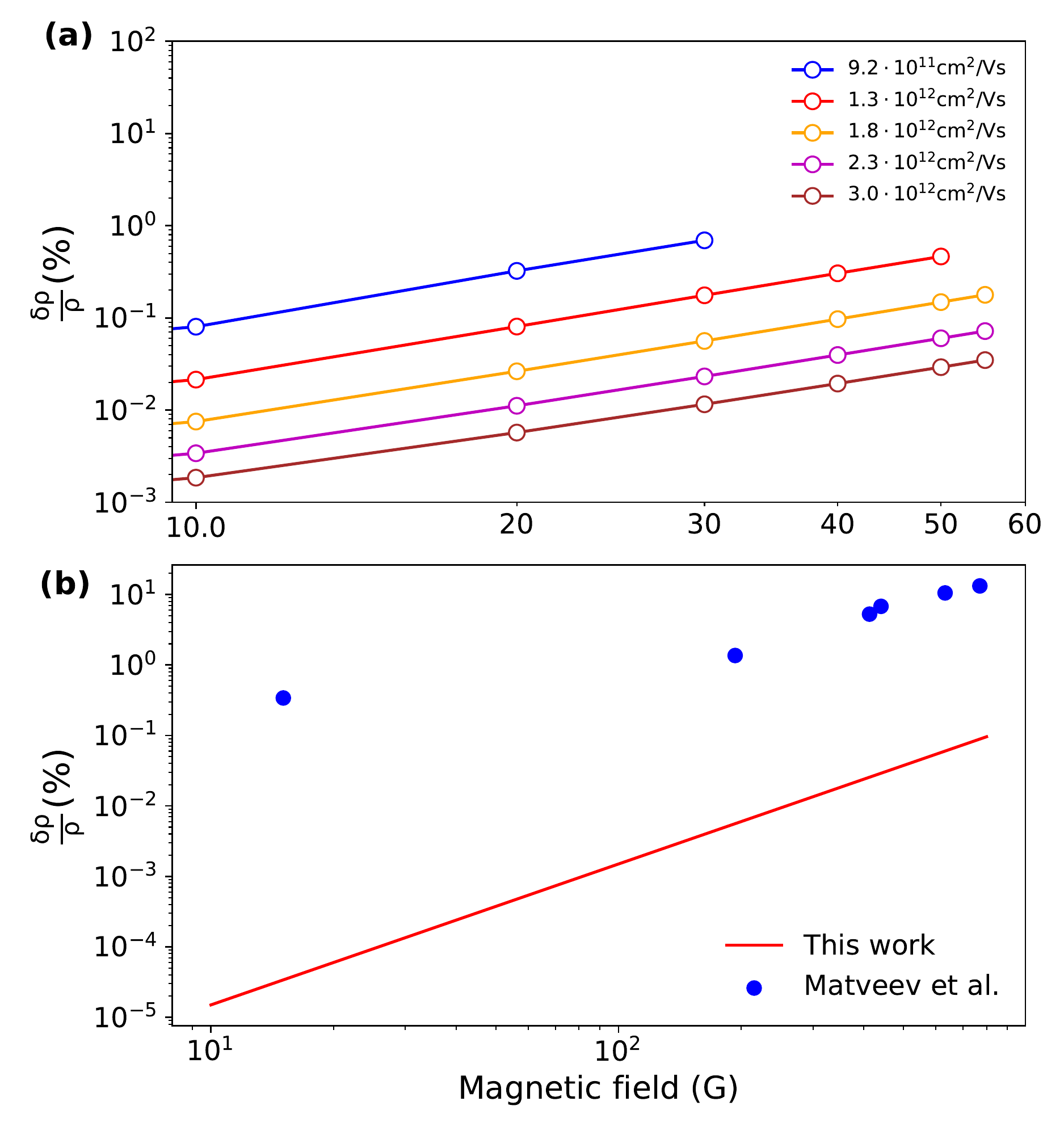}
\vspace{-15 pt}
\caption{(a) Transverse MR in graphene vs magnetic field for various hole carrier concentrations at 300~K. (b) Comparison of the MR to experimental data from Ref.~\cite{Matveev_Graphene_2018}. 
}
\label{fig:graphenemr}
\end{figure}

\vspace{1.5pt}
Similar to other semimetals~\cite{Ziman}, graphene exhibits a relatively large MR, with reported values of 20$-$50\% at room temperature and even greater at lower temperatures~\cite{Gopinadhan_Graphene_2013}. We discuss the MR in graphene for hole carriers but the MR values for electrons are similar. The accuracy of our settings is checked by calculating the drift mobility at 300~K; we obtain a value of $\sim$ $160000\,\, \mathrm{cm^2/Vs}$ consistent with experiments in suspended graphene~\cite{Bolotin_Graphene_2008}.
\\
\indent
Figure~\ref{fig:graphenemr}(a) shows the computed MR in graphene at 300~K. 
%
We find that the MR depends strongly on carrier concentration $-$ a doubling of concentration from $\sim$1.5 to $3\cdot 10^{12}\, \mathrm{cm}^2/\mathrm{Vs}$ decreases the MR by an order of magnitude. This situation makes comparison with experiment difficult [Fig.~\ref{fig:graphenemr}(b)] as the reported carrier concentration usually does not take into account the Hall factor (we find $r = 1.45$ for $n = 1.2 \cdot 10^{12}$ cm$^{-3}$, consistent with recent work~\cite{Macheda_Graphene_2020}). As by definition $n = r(e\, R_{\rm{H}})^{-1}$, carrier concentrations from Hall measurements are inaccurate unless the Hall factor is taken into account. In addition, most graphene samples are measured on substrates, often causing a reduction in the mobility. Accordingly, experimental values of the mobility and MR vary over a wide range~\cite{Gopinadhan_Graphene_2013,Wang_Graphene_2014,Rein_Graphene_2015}. This variability in the experimental results can at least partially explain the discrepancy between the calculated and measured MR in Fig.~\ref{fig:graphenemr}(b).
\\
\indent
Analysis of the electron occupations (see below) reveals that taking into account backscattering by iteratively solving the full BTE (as opposed to using the RTA) is essential in graphene, and that the RTA fails to capture the correct electronic occupations at steady state. This key role of backscattering in magnetotransport mimics trends found for thermal transport in graphene~\cite{Fugallo}.
\vspace{-9 pt}
%

%
\begin{figure*}[]
\centering 
\includegraphics[width=0.9\textwidth]{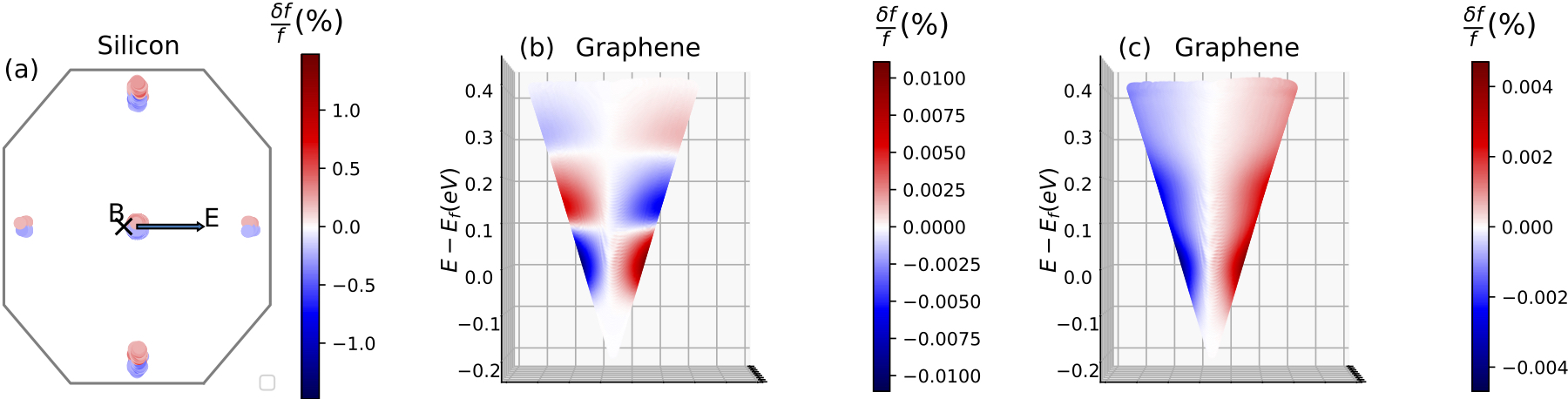}
\vspace{-5 pt}
\caption{(a) Projection of the occupation changes $\delta f / f$ onto the $k_z = 0$ plane in Si. 
(b) Occupation change $\delta f / f $ in graphene near the Dirac cone with contributions from all phonons. 
(c) The same result as in (b) for graphene but with scattering from the highest-energy optical phonon branch excluded from the transport calculation.}
\label{fig:grapheneocc}
\end{figure*}

\subsection{Steady-state occupations}
To conclude our analysis, we study the electron occupations at steady state, focusing on their change due to the magnetic field for a constant electric field. We define this relative occupation change as
$\delta f / f \!=\! [f_{n\mathbf{k}}(\mathbf{E},\mathbf{B}) - f_{n\mathbf{k}}(\mathbf{E},0)]/ f_{n\mathbf{k}}(\mathbf{E},0)$, and plot it in momentum space for Si and graphene. In the results for Si, shown in Fig.~\ref{fig:grapheneocc}(a), the occupation change projected on the $k_{z}$=0 plane clearly shows the effect of the Lorentz force, whereby the electrons deviate in the $\mathbf{E} \cross \mathbf{B}$ direction near the six conduction band minima (the occupations at the zone center are projections of the two band minima along the $k_{z}$ axis). As expected, the electrons are deflected in momentum space due to the magnetic field, an important sanity check for our numerical implementation.
\\
\indent
The results for graphene, shown in Fig.~\ref{fig:grapheneocc}(b), are more interesting. Similar to Si, the occupations near the Dirac cone are also changed by the Lorentz force. However, electrons in graphene couple strongly with LO phonons with momentum near $\Gamma$ and TO phonons with momentum near $K$ ~\cite{Xiao_2020}, which mediate intra- and inter-valley electronic processes respectively. As a result, optical phonon absorption generates a step-like pattern in the occupation changes, with 160$-$200 meV spacing equal to the LO and TO phonon energies~\cite{Xiao_2020}. 
The disappearance of the alternating patches on removing scattering from the highest optical branch from the transport calculation [Fig.~\ref{fig:grapheneocc}(c)] provides concrete evidence for the dominant optical phonon backscattering in graphene. The RTA completely misses this trend and gives occupation changes with a pattern similar to Fig.~\ref{fig:grapheneocc}(c). While in graphene the magnetotransport RTA results are in fairly good agreement with the full solution of the BTE, which correctly includes backscattering, our results show that this agreement has to be coincidental or due to error cancellation.
\\
\vspace{30pt}
\newpage
\section{CONCLUSION}
We have shown calculations of magnetotransport that can accurately predict the Hall mobility, Hall factor and MR in Si and GaAs. Our results for graphene leave room for improvements and call for stricter protocols for magnetotransport measurements in 2D materials. Analysis of the steady state occupations in graphene highlights a key strength of first-principles calculations $-$ they can capture the competition between mode-dependent $e$-ph scattering and the effect of the Lorentz force in momentum space, shedding light on the microscopic mechanisms governing magnetotransport. 
With calculations on materials with tens of atoms in the unit cell readily available~\cite{perturbo2020}, extension of these results to other semiconductors and 2D materials appears straightforward. The current formalism can be easily extended to include the Berry curvature, for example to study magnetotransport in topological semimetals and shed light on the origin of their unconventional MR. The availability of first-principles magnetotransport calculations, to be released in our {\sc{Perturbo}} code in the near future, greatly expands the reach of first-principles transport studies, connecting them more deeply with transport experiments, which are often carried out in magnetic fields.\\
%
%
\section{ACKNOWLEDGEMENTS}
This work was supported by the National Science Foundation under Grants No. DMR-1750613. J.-J.Z. acknowledges partial support from the Joint Center for Artificial Photosynthesis, a DOE Energy Innovation Hub, as follows: the development of some computational methods employed in this work was supported through the Office of Science of the US Department of Energy under Award No. DE-SC0004993. This research used resources of the National Energy Research Scientific Computing Center, a DOE Office of Science User Facility supported by the Office of Science of the US Department of Energy under Contract No. DE-AC02-05CH11231.

\end{document}